\documentclass{jfm}
\usepackage{graphicx}
\usepackage{graphicx}
\usepackage{epstopdf, epsfig}
\usepackage{amsmath}
\usepackage{color}

\usepackage{sistyle}
\shorttitle{Free-fall velocities in liquid metal magneto-convection}
\shortauthor{T. Vogt, J. Yang, F. Schindler, S. Eckert }

\usepackage{overpic}
\usepackage{rotating}

\usepackage[dvipsnames]{xcolor}
\newcommand{\TV}[1]{{\color{black}{#1}}}

\title{\TV{Free-fall velocities and heat transport enhancement in liquid metal magneto-convection}}
\author{Tobias Vogt\aff{1,2}
  \corresp{\email{t.vogt@hzdr.de}},
  Juancheng Yang\aff{1,2}
	 Felix Schindler\aff{1}
 \and Sven Eckert\aff{1}}

\affiliation{\aff{1}Institute of Fluid Dynamics, Helmholtz-Zentrum Dresden-Rossendorf, 01328 Dresden, Germany
\aff{2}State Key Laboratory for Strength and Vibration of Mechanical Structures, School of Aerospace, Xi'an Jiaotong University, Xi'an, Shaanxi 710049, P.R. China}

\begin{document}
\maketitle

\begin{abstract}
In geo- and astrophysics, low Prandtl number convective flows often interact with magnetic fields. Although a static magnetic field acts as a stabilizing force on such flow fields, we find that self-organized convective flow structures reach an optimal state where the heat transport significantly increases and convective velocities reach the theoretical free-fall limit, i.e. the maximum possible velocity a fluid parcel can achieve when its potential buoyant energy is fully converted into kinetic energy. Our measurements show that the application of a static magnetic field leads to an anisotropic, highly ordered flow structure and a decrease of the turbulent fluctuations. When the magnetic field strength is increased beyond the optimum state, Hartmann braking becomes dominant and leads to a reduction of the heat and momentum transport. The results are relevant for the understanding of magneto-hydrodynamic convective flows in planetary cores and stellar interiors in regions with strong toroidal magnetic fields oriented perpendicular 
to temperature gradients.
\end{abstract}

\section{Introduction}
\label{sec:intro}
\TV{Turbulent convective energy coalesces into large coherent flow structures. This is one of
the key features that delineates many geo- and astrophysical systems, and also manifests
in numerous industrial applications.} Often, these systems are subjected to additional, stabilizing forces such as centrifugal and Coriolis forces due to rotation, or Lorentz forces due to magnetic fields. These forces can substantially affect the flow structures and therefore, the heat and momentum transport of convective systems.

The simplest physical model to study thermally driven flows is the so-called 
Rayleigh-B\'enard convection (RBC), where the driving force is a temperature gradient $\nabla \textbf{T}$ between a warmer bottom and 
a cooler top \citep{ahlers2009, chilla2012}. An important outcome of RBC studies are the scaling relations for the global 
heat and momentum transport, expressed non-dimensionally in terms of Nusselt number ($Nu$) and Reynolds number 
($Re$), respectively \citep{grossmann2002, stevens2013}.

When stabilizing forces such as rotation \citep{stellmach2014, guervilly2014}, geometrical confinements \citep{daya2001, huang2013} or static magnetic fields are added to highly non-linear systems such as RBC, unexpected features are encountered \citep{chong2017, aurnou2018}. For instance, studies have shown that rotation 
around a vertical axis has strong influence on the flow structure and consequently the heat transport. 
Although it is well known that the Coriolis force has a stabilizing effect \citep{proudman1916, taylor1917, chandrasekhar1961}, it was found that the application of moderate forces can even enhance scalar transport \citep{rossby1969, zhong1993, liu2009, stevens2009, wei2015, weiss2016, chong2017}.

Similarly, magnetic fields exert an influence on electrically conducting fluids by the induction of eddy currents 
$\textbf{j} = \sigma (\textbf{E} + \textbf{u} \times \textbf{B}$) which give rise to a corresponding Lorentz force 
$\textbf{f}_l = \textbf{j} \times \textbf{B}$ 
that acts on the fluid. Here, $\sigma$ is the electrical conductivity, \textbf{E} is the electric field, $\textbf{u}$ is the velocity and $\textbf{B}$ is the magnetic 
field. A static magnetic field cannot generate any flow from a quiescent state, however, it can reorganize an electrically 
conducting flow field so as to minimise the Joule dissipation \citep{davidson1995}. This is a direct consequence of a reduction of the velocity gradients along the magnetic field direction
due to the induced eddy currents \citep{sommeria1982, potherat2017}.

In thermal convection in liquid metal, the orientation of the applied magnetic field with respect to the the temperature gradient $\nabla \textbf{T}$, plays a pivotal role on the details of the resulting flow field. Two field orientations are possible: vertical and horizontal.
A vertical magnetic field ($\textbf{B} \parallel \nabla \textbf{T}$) inhibits the onset of liquid metal convection and the heat transport decreases 
monotonically with increasing the magnetic field strength, due to a strong suppression of the bulk flow \citep{cioni2000, aurnou2001, burr2001vertical, liu2018, yan2019, zuerner2020}. \TV{Recently, it was shown that convection in fluid with $Pr=8$, and under the influence of a vertical magnetic field results in increased heat flux with increasing magnetic field strength but accompanied by a decrease in momentum transport \citep{lim2019}}.

A horizontal magnetic field ($\textbf{B} \perp \nabla \textbf{T}$) however, converts the convective motion into a flow pattern of quasi-two-dimensional 
rolls arranged parallel to the magnetic field \citep{fauve1981, busse1983, burr2002, yanagisawa2013, tasaka2016, vogt2018a}.

In this paper, we report about the interaction between liquid metal convection and a static, horizontally imposed magnetic field, \TV{which is governed by two non-dimensional parameters. The Rayleigh number 
\begin{equation}
	Ra=\alpha g \Delta T H^3 / \kappa \nu
\end{equation}
is a measure of the thermal forcing that drives the convection. The strength of the stabilizing Lorentz force due to the applied magnetic field is expressed by the Chandrasekhar number 
\begin{equation}
Q = B^2 L^2 \sigma /\rho \nu = Ha^2. 
\end{equation}
Here, $H$ is the distance between the heated and the cooled plates, $L$ is the width of the cell, $\Delta T$ is the imposed temperature difference between these plates, $B$ is the strength of the magnetic field, $\rho$ is the density of the liquid metal, $\alpha$ 
is the isobaric expansion coefficient, $\nu$ is the kinematic viscosity, $\kappa$ is the thermal conductivity, $g$ denotes the gravitational acceleration and $Ha$ is the Hartmann number.}

\TV{The present work is a continuation of the work of \cite{vogt2018a}, where the transition from a three-dimensional to a quasi two-dimensional flow structure in a liquid metal convection under the influence of a horizontal magnetic field was investigated. \cite{vogt2018a} focused on the qualitative description of large and small-scale flow structures at different parametric combinations of Rayleigh number and Chandrasekhar number. The main goal of this work is to investigate how this transition between flow regimes affects heat and momentum transport.
Accordingly, the measuring arrangement at the experiment was extended and the number of measurements was increased significantly to allow a fine increment of the Chandrasekhar number.}
We find that the reorganization of the convective flow due to the magnetic field
results in a significant enhancement of both heat and momentum transport. 
In the optimum, the convective velocities can even reach the free-fall limit $u_{f\!f}=\sqrt{\alpha g \Delta T H}$. In classical Rayleigh-B\'enard convection in fluids with moderate Prandtl numbers, such as water or air, the flow velocities are well below the free-fall limit and do not exceed $u_{max}/u_{f\!f} \leq 0.2$ \citep{niemela2003}. Therefore, our measurements demonstrate how intense low $Pr$ magnetohydrodynamic convective flows can actually become.\\
%

\section{Laboratory Magneto-Convection Experiments}\label{sec:setup}
\subsection{\TV{Experimental set-up}}
\begin{figure}
\centerline{\includegraphics[width=1\textwidth]{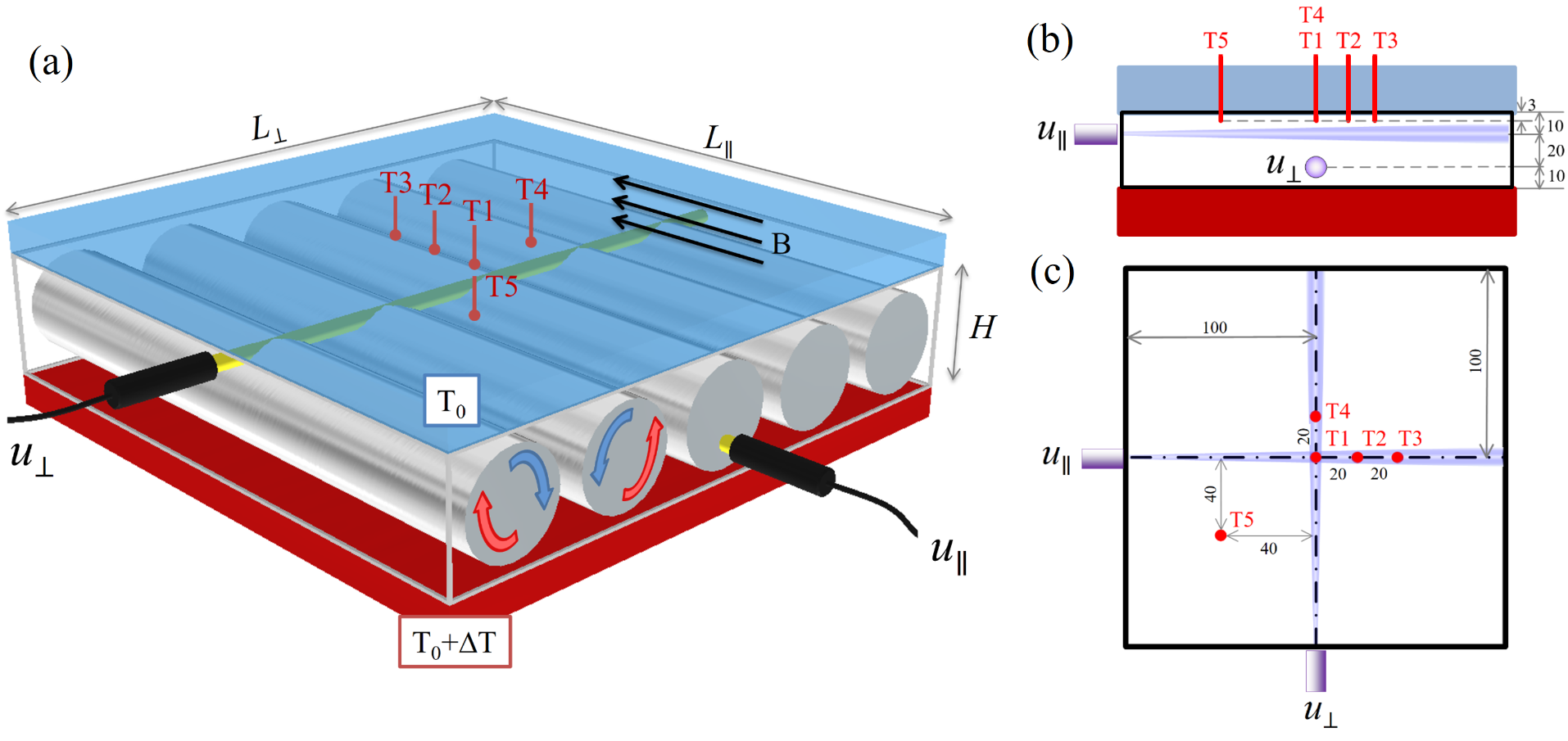}}
\caption{\TV{Schematic of the experimental set-up (a). The convection rolls (shown in grey), driven by the temperature difference $\Delta T$ between the heated bottom plate (red) and the cooled top (blue), are aligned parallel to the magnetic field $(B)$. The velocities $(u_{\parallel}, u_{\bot})$ were measured with ultrasound transducers (black tubes) parallel and perpendicular to the 
direction of the magnetic field. Five thermocouples $(T1-T5)$ measured the temperature  inside the fluid, at a distance of 3 mm from the inner edge of top plate. The side view (b) and the top view (c) schematics illustrate the positions for the ultrasound transducers and thermocouples. The units are in mm.}}
\label{fig_setup}
\end{figure}

\TV{The experiments were conducted at the Helmholtz-Zentrum Dresden Rossendorf (HZDR). The eutectic liquid metal alloy composed of Gallium, Indium, and Tin (GaInSn, melting point of $T=10.5^\circ\text{C}$, $Pr=\nu / \kappa = 0.03$) was used as the working fluid \citep{plevachuk2014}. The liquid metal is contained in an aspect ratio $\Gamma = L/H = 5$ rectangular vessel with a cross section
$L^2=200 \times 200$ $\mathrm{mm^2}$ and a height $H=40$ mm (Fig. \ref{fig_setup}). The side walls are made of 30 mm thick Polyvinyl chloride and the top and bottom are made of copper. The convection cell is wrapped in 30 mm closed-cell foam to minimize heat loss. The temperature of the top and bottom were adjusted by a constant temperature water bath which flows through channels in the copper plates. The maximum heat flux is 1500 Watts. The applied temperature drop between the plates ranges from $1.1^\circ\text{C} \leq \Delta T \leq 11.7^\circ\text{C}$ whereby the mean fluid temperature was kept constant at $T_m = 20 ^\circ C$. The Rayleigh number ranges between $2.3 \times 10^4 \leq Ra \leq2.6 \times 10^5$. A static, uniform horizontal magnetic field penetrates the liquid metal with a strength $0 \leq B \leq 317$ mT, which gives a Chandrasekhar number range $0 \leq Q \leq 6.1 \times 10^6$.}

\TV{The fluid velocities are measured using Ultrasound Doppler Velocimetry (UDV), which provides instantaneous velocity 
profiles along two horizontal directions, as shown in figure \ref{fig_setup}. This technique is useful for non-invasively measuring velocities in opaque fluids \citep{brito2001, tsuji2005, vogt2014, vogt2020}. The transducers (TR0805SS, Signal Processing SA) detect the velocity component parallel to the ultrasonic beam with resolutions of about 1 mm in beam direction and 1 Hz in time. One UDV transducer measures the flow velocities perpendicular to the magnetic field ($u_{\bot}$) and is located in the middle of the cell width at $L_{\parallel}/2$ and $10$ mm below the upper boundary ($3H/4$). A second UDV transducer measures the magnetic field parallel velocity component ($u_{\parallel}$) and is also located in the middle of the cell width at $L_{\bot}/2$, 
but at a different height, $10$ mm above the lower boundary ($H/4$). Both transducers are in direct contact with the liquid metal which allows a good velocity signal quality even at low velocities.}

\TV{The difference between the mean temperatures of the heated and the cooled plates were obtained from two sets of nine thermocouples, with each set located in the heated and the cooled plates respectively. The thermocouples are individually calibrated using a high precision  thermometer to ensure accuracy better than 0.05 K.}

\TV{Another five thermocouples measure the temperature inside the liquid metal at a distance of 3 mm below the cold plate (c.f. figure \ref{fig_setup}(b,c))}

\TV{The convective heat transport is expressed dimensionless by means of the Nusselt number, $Nu=\dot{\Phi}/\dot{\Phi}_{cond}$. Here, $\dot{\Phi}_{cond}=\lambda L^2 \Delta T / H$ is the conductive heat flux, with $\lambda$ being the thermal conductivity of the liquid metal. $\dot{\Phi}=\rho c_p \dot{V} (T_{in}-T_{out})$ is the total heat flux injected at the bottom and removed from the top wall heat exchanger, whereby $c_p$ is the isobaric heat capacity of water. The total heat flux is determined by the flow rate $\dot{V}$ and the temperature change $(T_{in}-T_{out})$ of the circulating water inside the hot and cold wall heat exchangers.}

\subsection{\TV{Non-dimensional quantities and characteristic length scale}}

\TV{The length, velocity and time are made non-dimensional throughout this work using the 
width $L$ of the cell, the free-fall velocity $u_{f\!f}$, and the free-fall time 
$t_{f\!f}=H/u_{f\!f}$, respectively.}

\TV{In contrast to the work of Vogt (2018) the Chandrasekhar number was not determined with the cell height $H$ but with the distance $L_{\parallel}$ of the horizontal walls in magnetic field direction. The definition of $Q$ using the height $H$ goes back to the studies of \cite{burr2002}, who used the same definition of the Hartmann number $Ha$ with $H$
as characteristic length for their investigations both in the vertical \citep{burr2001vertical} and the horizontal magnetic field \citep{burr2002}. In our opinion, the use of the horizontal dimension of the cell is better suited for the case of a horizontal magnetic field, since the effect of Hartmann braking scales with the dimension of the flow domain in
magnetic field direction \citep{mueller2001, knaepen2008}.}

\begin{figure}
\centerline{\includegraphics[width=0.9\textwidth]{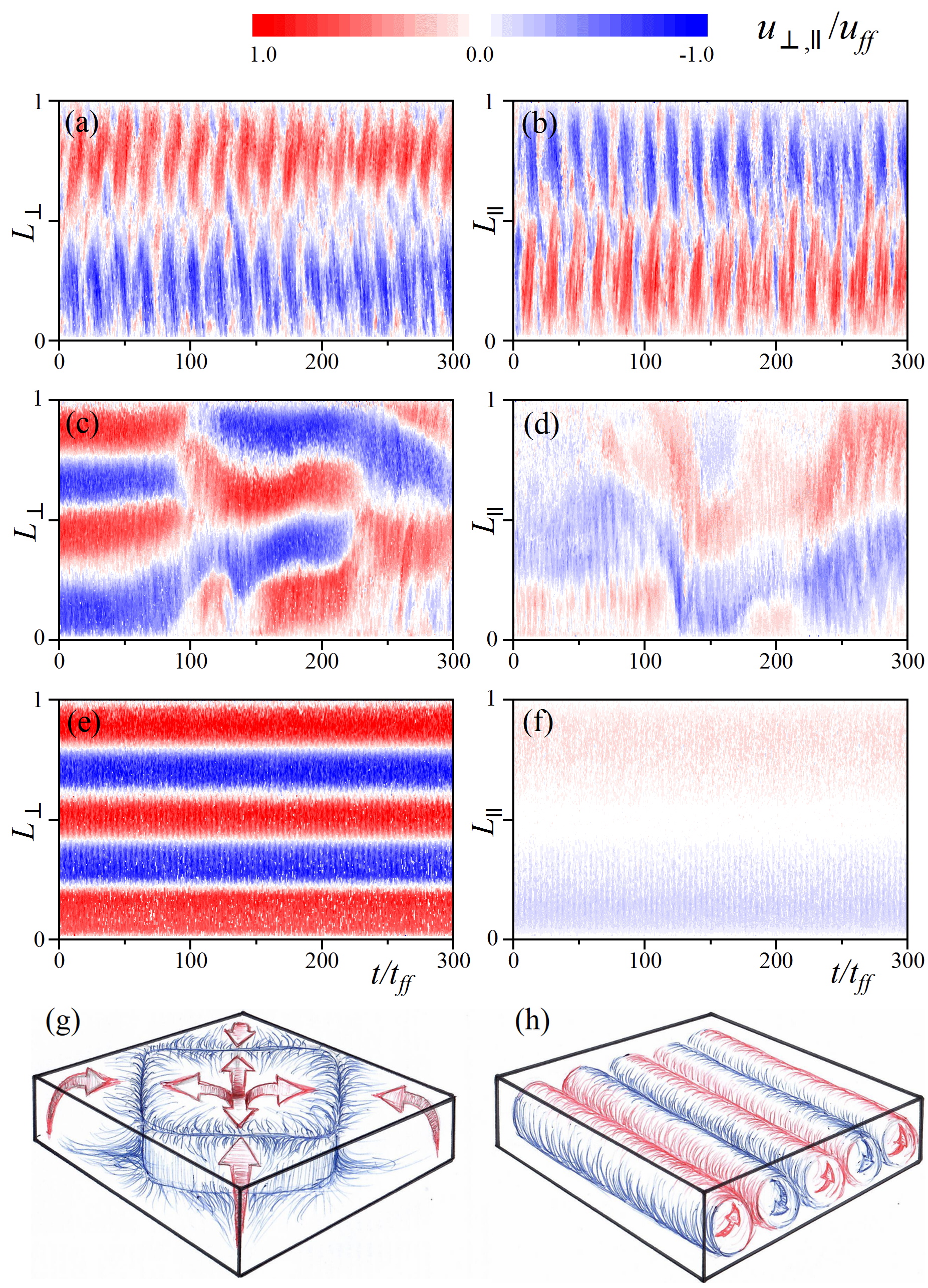}}
\caption{UDV Dopplergrams: Spatiotemporal distribution of the horizontal velocity measured perpendicular (a,c,e) 
and parallel (b,d,f) to the magnetic field direction at $Ra=2.18 \times 10^5$. The measuring lines referred to are indicated in Fig. 1. A positive (negative) velocity 
represents a flow away from (towards) the transducer. The measured velocities are non-dimensionalised 
using the free-fall velocity $u_{f\!f}=\sqrt{\alpha g \Delta T H} = 21.9$ mm/s and the free-fall 
time $t_{f\!f}=H/u_{f\!f}=1.8$ s. The ordinate corresponds to the measuring depth along the 
horizontal dimensions $L_{\bot}$ and $L_{\parallel}$ of the container. An increase of the 
magnetic field strength changes the global flow structure from: (a,b) oscillating cell 
structure at $Q=529$, (c,d) unstable 3,4,5-roll configuration at $Q=8.7 \times 10^4$, (e,f) stable 5-roll 
configuration at $Q=3.8 \times 10^5$. (g) Schematic illustration of the cell regime and (h) the stable 5-role regime. 
Red areas symbolize warm, ascending fluid and blue areas symbolize colder, descending fluid.}
\label{fig_udv}
\end{figure}

\section{Results}\label{sec:results}
\TV{All results presented here were recorded after the temperature
difference between the hot and the cold plates reached a constant value, and the system had attained thermal equilibrium.
At low magnetic field strength,} the convection at sufficiently high $Ra$ forms a 
large scale circulation with a three-dimensional cellular structure that fills the 
entire cell (Fig. \ref{fig_udv}a,b,g), whereby upwelling takes place in the center and 
all four corners of the vessel. A detailed description of this structure can be found 
in \citep{akashi2019}. The large amplitude oscillation that can be seen in 
Fig. \ref{fig_udv}(a,b) is a typical feature of inertia-dominated liquid metal flows 
due to their low viscosity and high density \citep{vogt2018b}. Applying a horizontal magnetic field 
to such a three-dimensional flow promotes the formation of quasi two-dimensional convection 
rolls that are aligned parallel to the magnetic field lines. The number of rolls formed
depends on the ratio between the driving and the stabilizing force $Ra/Q$ and the aspect ratio $\Gamma$ of the vessel \citep{tasaka2016}. 
Fig. \ref{fig_udv}(c,d) shows the example of a convective flow field under the influence of an intermediate 
magnetic field strength and exhibits an unstable roll configuration. \TV{In this range the magnetic field is not yet intense enough to produce a stable quasi-two-dimensional flow in perfection. The character of the global flow is still three-dimensional, but the magnetic field has caused a breaking of the symmetry, which characterizes the cell structure. The flow structure is dominated by the convection rolls, but their shape and orientation is still transient and subject to strong three-dimensional disturbances.} Four rolls are formed in this transitional range, but these are irregular, and temporary changes to three or five roll configurations can occur. At higher 
field strength, the flow develops five counter rotating convection rolls which are 
very stable in time (Fig.\ref{fig_udv}e,f,h). At this stage, convection has restructured into a quasi two-dimensional flow field 
oriented parallel to the magnetic field direction. The symmetric but weak flow that appears along $L_{\parallel}$ 
is evoked by the Ekman-pumping that originates in the B\"odewadt boundary layers where 
the convection rolls meet the sidewalls \citep{vogt2018a}. \TV{A weak but regular oscillation is visible in figure \ref{fig_udv}(e,f), which is due to inertial waves within the convection rolls \cite{yang2020}. However, apart from these regular oscillations, the flow appears to be laminar.}\\
Based on the flow fields shown in Fig. \ref{fig_udv}, we distinguish here mainly between three characteristic regimes, 
the "cell structure" where the influence of the magnetic field on the flow is 
negligible, the "unstable 3,4,5-roll" state where the field starts to reorganize 
the flow but is not strong enough to form stable roll configurations, and finally the "stable 5-roll" state at higher Chandrasekhar numbers that results in the formation of stable, quasi two-dimensional convection rolls.\\
\begin{figure}
\centerline{\includegraphics[width=0.7\columnwidth]{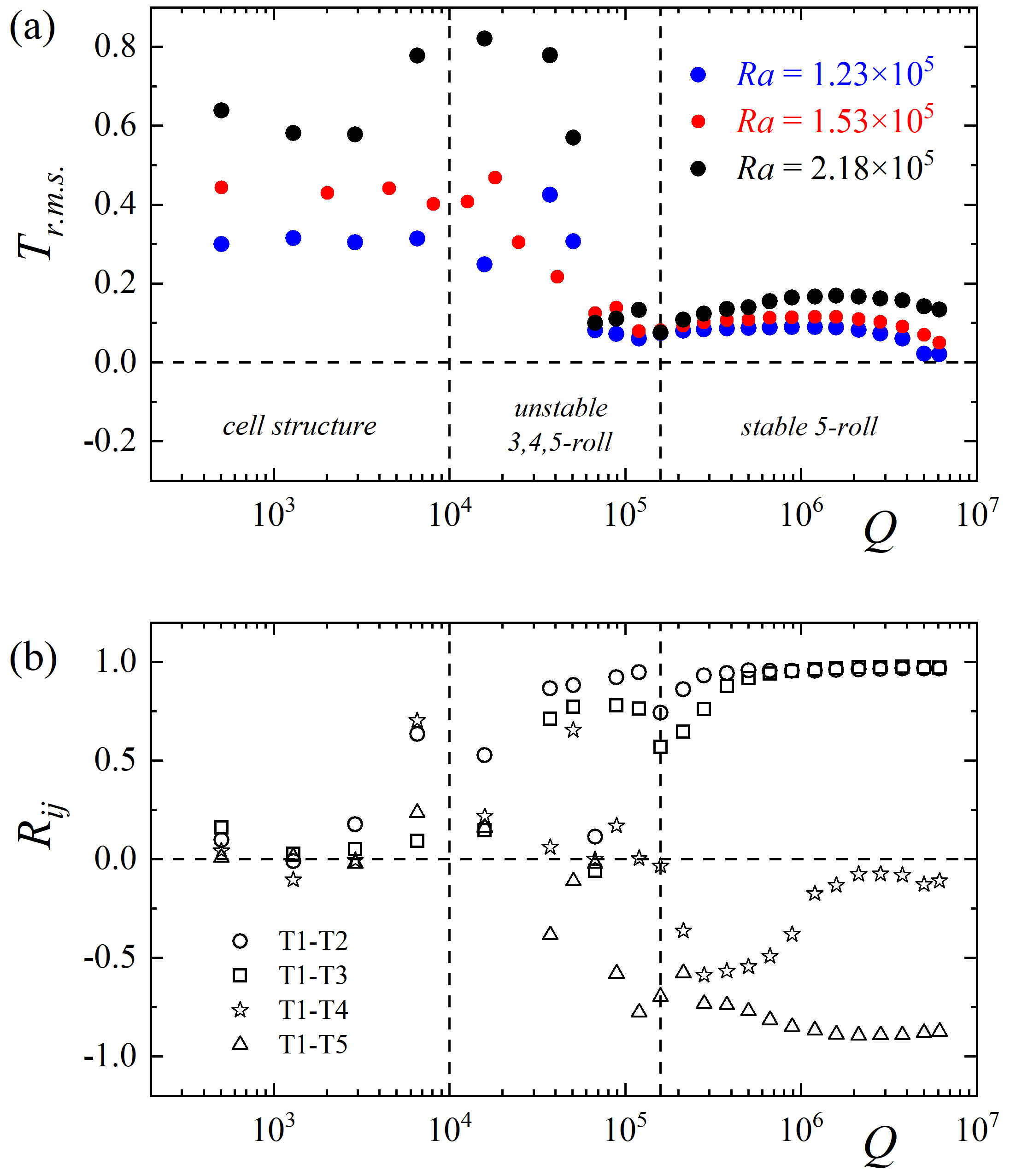}}
\caption{
\TV{Liquid metal temperature measurements in a 3 mm distance from the top plate and their dependence on $Q$. (a) Averaged r.m.s. of temperature fluctuation of T1 for three different $Ra$ numbers. (b) Averaged cross-correlation coefficient at different temperature measuring positions at $Ra = 2.18 \times 10^5$. The different symbols denote the value of cross-correlation coefficients between corresponding thermocouples: T1 and T2 (circles), T1 and T3 (squares), T1 and T4 (stars), and T1 and T5 (triangles)}
\label{fig_temp}}
\end{figure}
\TV{Figure \ref{fig_temp}(a) shows the averaged root-mean-square (r.m.s.) of temperature fluctuations $T_{r.m.s.}$ for all three regimes measured with thermocouple T1 which is under the cold plate, dipped 3 mm into the liquid metal at the cell centre. The temperature fluctuations are calculated as
\begin{equation}
	T_{r.m.s.} 	= \sqrt{\frac{\sum_{t=0}^{t_{end}}(T(t)-T_{avg})^{2}}{N}}
\end{equation}
with $T_{avg} 	= <T(t)>_{t}$ the average over the whole measurement and $N$ the number of measurement points. The vertical dashed lines in Figure \ref{fig_temp}(b) show the boundaries between the three different flow regimes. However, this is only indicative since the actual regime boundaries depend not only on $Q$ but also on $Ra$. The fluctuations are strongest in the cell structure regime and the unstable 3,4,5-roll regime. The larger the $Ra$ number, the stronger the fluctuations. At the transition to the stable 5-roll regime, the fluctuations decrease significantly and are close to zero, which indicates that from this point on the position and orientation of the rolls within the convection cell is arrested by the applied magnetic field. The slight but systematic increase of the fluctuations in the stable regime is surprising at first sight, but can be explained by the occurence of inertial waves inside the convection rolls \citep{yang2020}. Finally, at very high magnetic field strengths, these oscillations are also damped and the temperature fluctuations decrease again and approach zero.}
\TV{Figure \ref{fig_temp}(b) shows the cross correlation of the different temperature sensors within the liquid metal calculated as
\begin{equation}
	R_{i,j} 	= \frac{\sum_{t=0}^{t_{end}}(T_{i}(t)-T_{i,avg})(T_{j}(t)-T_{j,avg})}{\sqrt{\sum_{t=0}^{t_{end}}(T_{i}(t)-T_{i,avg})}\sqrt{\sum_{t=0}^{t_{end}}(T_{j}(t)-T_{j,avg})}}
\end{equation}
whereby $R_{i,j}$ is the Pearson's correlation coefficient. In the cell structure regime all cross correlation coefficients are scattered around zero and indicate a negligible correlation between the different measurement points due to a complex and turbulent flow field. In the unstable roll regime, first rolls form along the magnetic field and the cross correlation coefficient between the corresponding adjacent sensors in the magnetic field direction increases and approaches $R_{i,j} \approx 1$.

In the stable 5-roll regime, sensors T1, T2 and T3 are located along the same roll. The sensor T5 is located centrally above the neighboring convection roll with opposite rotation direction. Sensor T4 is located centrally between two neighboring convection rolls.\\ 
During transition to the stable 5-rolle regime, the values for $R_{T1-T2}$ and $R_{T1-T3}$ decrease initially, and then approache a value of 1 again. Correlation of about 1 is expected as these thermocouples (T1, T2 and T3) are located along the same roll. The reason for the initial decrease of the cross correlation coefficient is that the oscillations at the beginning of the stable roll regime are still weak and three-dimensional in nature \citep{yang2020}. With increasing magnetic field strength the three-dimensional character of the oscillations is suppressed, and from $Q>10^6$ onwards only quasi-two-dimensional oscillations take place. The cross correlation coefficient of $R_{T1-T2}$ and $R_{T1-T3}$ then reaches its maximum. The oscillations of neighboring rolls take place with a phase shift of $\pi$. For this reason, T5 which is located centrally above the neighbouring roll with opposite direction of rotation, registers a cross correlation coefficient $R_{T1-T5} \approx -1$. At the measuring position T4 between two adjacent rolls, remaining oscillations vanish with increasing $Q$ and the correlation coefficient $R_{T1-T4}$ approaches zero for high magnetic field strengths.}

\begin{figure}
\centerline{\includegraphics[width=0.7\columnwidth]{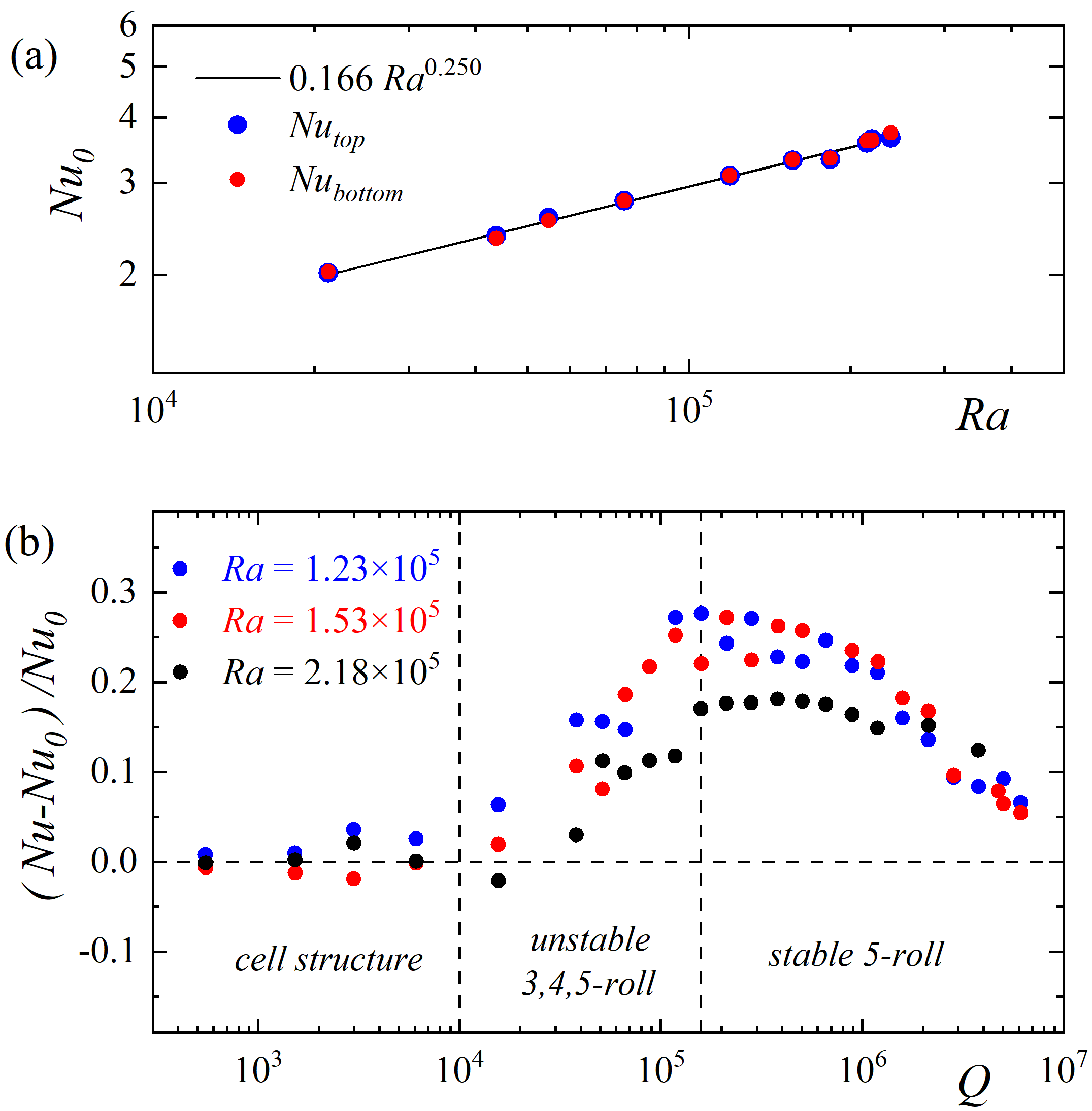}}
\caption{
\TV{(a) Measured Nusselt number $Nu_{\textit{0}}$ for convection without magnetic field. $Nu_{top}$ and $Nu_{bottom}$ are based on the total heat flux measured at the top and bottom heat exchanger, respectively. (b)} Relative deviations of the heat transfer from the reference state of RBC (without magnetic field), ($Nu-Nu_{\textit{0}})/Nu_{\textit{0}}$
as a function of $Q$ for three different $Ra$ numbers. The heat transfer reaches its maximum 
at $Q\approx 2.5  \times 10^5$ when the magnetic field forms stable convection rolls aligned parallel to the magnetic field.
\label{fig_nu}}
\end{figure}

For the case of RBC without magnetic field, our measurements show that the heat transport properties scale as: $Nu_{\textit{0}} = 0.166 Ra^{0.250}$ as shown in figure \ref{fig_nu}(a). This is in reasonable agreement to $Nu_{\textit{0}} = 0.147 Ra^{0.257}$ measured in Mercury \citep{rossby1969} 
and $Nu_{\textit{0}} = 0.19 Ra^{0.249}$ measured in Gallium \citep{king2013}. Note, that Mercury, Gallium 
and GaInSn have comparable Prandtl numbers ranging from $Pr=0.025-0.033$. Fig. \ref{fig_nu}(b) 
presents the relative deviation ($Nu - Nu_{\textit{0}})/Nu_{\textit{0}}$ for convection with imposed magnetic 
field where $Nu_0$ is the corresponding Nusselt number for RBC (without magnetic field). In the case of cellular flow structures, which is the prevailing structure 
for $0<Q<1 \times 10^4$, the heat transfer does not vary remarkably with increasing $Q$. 
This behavior changes in the range $1 \times 10^4 <Q< 1.6 \times 10^5$ where the formation of unstable convection rolls proceeds with significant increase of heat transfer. Finally, for $Q>1.6  \times 10^5$, $Nu$  reaches a maximum 
before the heat transfer decreases for even higher $Q$. The investigation of a wider $Ra$ range 
is not possible with our current experimental setup. \TV{This is due to limited power range of the thermostats, and the cell height which limits Rayleigh number}, $Ra_{max} \approx 3 \times 10^5$. On the other hand, lowering $Ra \leq 10^5$ triggers a transition from a stable 5-roll to a 4-roll structure \TV{regime which is beyond the scope of this paper, as we focus only on} the $Ra$ range where the stable 5-roll configuration fits well into the aspect ratio $\Gamma = 5$ of the cell.
\begin{figure}
\centerline{\includegraphics[width=0.8\columnwidth]{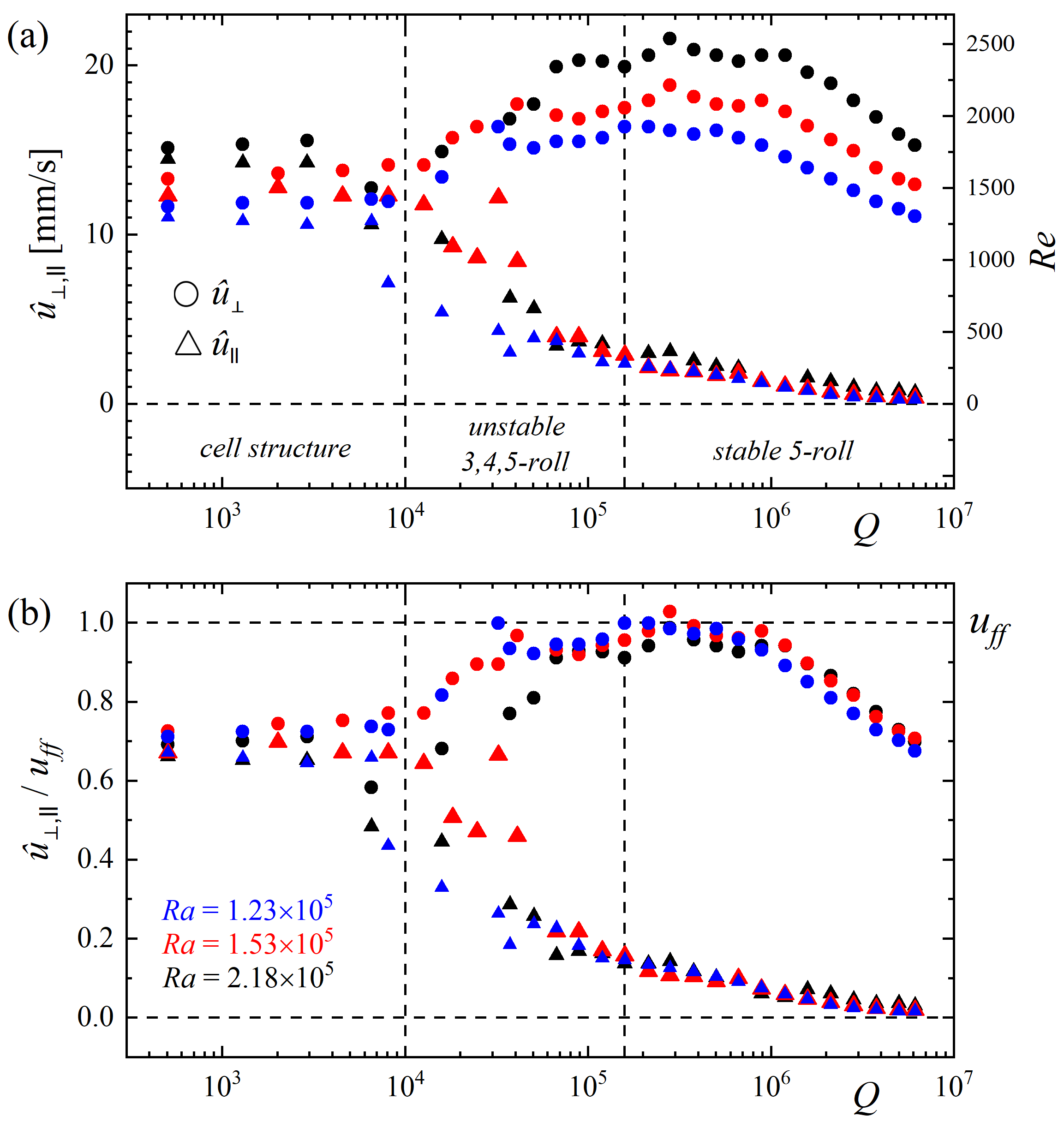}}
\caption{
\TV{(a) $Q$ dependence of the $\hat{u}_{\bot}$ and $\hat{u}_{\parallel}$ velocity components and the corresponding Reynolds number $Re = \hat{u}_{i} H / \nu$. (b)} $Q$ dependence of the normalized horizontal velocity amplitudes for three different $Ra$ numbers. 
The field-normal velocity amplitude increases from $\hat{u}_{\bot}/u_{f\!f}\approx 0.7$ at cellular flow structures ($Q<1 \times 10^4$) 
to $\mathcal{O}(u_{f\!f})$ in the stable 5-roll regime ($Q\approx 2.5 \times 10^5$). The diverging branches of $\hat{u}_{\parallel}$
and $\hat{u}_{\bot}$ start with the transition from cellular flow structure to magnetic field aligned convection rolls at $Q>1 \times 10^4$.
\label{fig_velo}}
\end{figure}
Enhancement of heat transfer in a liquid metal layer due to the application 
of a horizontal magnetic field was also investigated by \cite{burr2002}. Temperature 
measurements revealed an increase of $Nu$ in a certain range of $Q$. The correlation 
of temperature signals suggests that the enhancement of the convective heat transfer 
is accompanied by the existence of non-isotropic time-dependent flows. However, there are 
still no direct flow measurements of this phenomenon to explain the increase in convective 
heat transport.\\

Based on velocity measurements, we analyze the $Q$ dependance of the amplitude of the velocity components perpendicular
$\hat{u}_{\bot}$ and parallel $\hat{u}_{\parallel}$ to the magnetic field as shown in Fig. \ref{fig_velo}. The maximum velocity values $\hat{u}_i$ for figure \ref{fig_velo} were determined as follows:
For each measurement, a velocity threshold was defined, such that 95$\%$ of the 
velocity values of a measurement are below the threshold value. This approach provides very 
reliable values for the vast majority of measurements. Only at the largest $Q$ and the associated 
very low velocities $\hat{u}_{\parallel}$, the signal-to-noise ratio of the velocity measurements 
is not sufficient for applying this method. For these measurements, $\hat{u}_{\parallel}$ was determined from 
the time-averaged quasi stationary velocity profile.

\TV{The flow velocities, and as such the $Re$ number increases with $Ra$ in all three regimes.}
As for the heat transfer, the velocity components of the cell structure do not significantly 
change for $Q<1 \times 10^4$. Both velocity components, $\hat{u}_{\bot}$ and $\hat{u}_{\parallel}$ are at the 
same order of magnitude and reach an amplitude of about $\hat{u}_{\bot}/u_{f\!f} \approx 0.7$, which is an expected velocity value for 
low $Pr$ thermal convection at this $Ra$ number range \citep{vogt2018b, zuerner2019}. 
For $Q>1 \times 10^4$, the development of the unstable 3,4,5-roll state goes along with a separation 
of the velocity components. The increase of $\hat{u}_{\bot}$ and the decrease of $\hat{u}_{\parallel}$ 
indicates that the flow field starts to become quasi two-dimensional. 
The relatively large scatter of the velocity data in this regime is caused by the transient flow behavior with frequent reversals of the flow direction.
At $Q > 1.6 \times 10^5$ the flow structure changes into the stable 5-roll state, which remains 
the dominant flow structure for at least one decade of $Q$ numbers. The small scattering 
of the measured velocity amplitudes in this regime reflects the stable 
characteristic of this flow configuration. The velocity component parallel to the magnetic field $\hat{u}_{\parallel}$ decreases monotonically for higher $Q$ while $\hat{u}_{\bot}$ reaches maximum arround $Q \approx 2.5 \times 10^5$, where the velocity amplitudes reach the theoretical free-fall limit $u_{f\!f}$. 
The normalization of the velocity amplitude with the free-fall velocity yields good conformity 
for the different $Ra$ numbers. 

\section{Discussion}

We have demonstrated that the rearrangement of a three-dimensional 
thermal convection flow into a quasi two-dimensional flow field, due to an applied static magnetic field, results in significantly increased heat and momentum transport. The convection 
forms five counter rotating rolls, whereby the diameter of the rolls corresponds to the height of the fluid layer and the number of rolls results from the aspect ratio $\Gamma = 5$ of the vessel. 
The preferred orientation of the rolls implies that momentum 
oriented parallel to the magnetic field is redirected in the direction perpendicular to the field. Therefore, $\hat{u}_{\parallel}$ decreases while $\hat{u}_{\bot}$ increases. In addition, the intensity of fluctuations in the temperature and velocity field decreases and the stabilized 
convection rolls appear laminar and quasi-stationary.

\TV{The vertical velocity component $u_z$ is responsible for convective heat flux, but this component was not directly measured in the experiment. However, our measurements show a fully three-dimensional flow in the cell structure regime wherein the three velocity components are of similar amplitude. It can therefore be assumed that the velocity components in this regime are as follows:}: $u_z \approx u_{\bot} \approx u_{\parallel} \approx 0.7 u_{f\!f}$. By contrast, in the quasi two-dimensional, stable 5-roll regime, the flow component parallel to the magnetic field \TV{was measured to be} significantly weaker compared to the flow component perpendicular to the magnetic field. \TV{These experimental results, in conjunction with the prevailing topology of the flow structure, and the requirement imposed by continuity indicate that the following relation would hold for the velocity components at} the optimal state of the stable 5-roll regime: $u_z \approx u_{\bot} \approx u_{f\!f}$.\\

\TV{In classical Rayleigh-B\'enard convection in fluids with moderate Prandtl numbers, such as water or air, the flow velocities are well below the free-fall limit and never exceed $u_{max}/u_{f\!f} \leq 0.2$ \citep{niemela2003}. Our results show unequivocally that the flow has a predisposition to reorient itself perpendicular to the magnetic field, which allows it to attain the optimal state wherein the fluid parcel traverses with the maximum possible velocity, the free-fall velocity. Consequently, the vigour of the convective flow in such state is intense, leading to an enhancement of the heat flux. Further increase of Lorentz force or $Q$, beyond the optimum state} yields reduced convective transport due to the increasing dominance of the Hartmann braking in the lateral boundary layers perpendicular to the magnetic field \citep{vogt2018a, yang2020}. \TV{ In an infinite fluid layer, the Hartmann braking for an ideal two-dimensional flow structure aligned with the magnetic field direction would disappear} since the characteristic Hartmann damping time scale $\tau_{H\!B} = \sqrt{{\rho L^2}/{\sigma \nu B^2}}$ shows a linear dependence on the distance $L$ between the Hartmann walls \citep{sommeria1982}.\\
\TV{In previous works \citep{chong2017,lim2019}, the increase in Nusselt number was explained as a result of increased coherency of the flow structures, which acts} as the main carrier for the heat transport. Moreover, the authors concluded that the \TV{maximum} heat flux is achieved when the thermal and viscous boundary layers reach the same thickness. Our results differ in several respects from the studies mentioned above. First, the low $Pr \ll 1$ implies that the viscous boundary layer is always nested well inside the much thicker thermal boundary layer. A crossover of the boundary layer thicknesses is therefore, not expected in very low Prandtl number fluids such as liquid metals. Second, in our case, not only the heat flux, but also the momentum transport perpendicular to the magnetic field is increased. And finally, the low $Pr$ of liquid metals implies that the Peclet number $Pe = Re \, Pr$ remains low when compared to moderate $Pr$ flows at a comparable turbulence level \citep{vogt2018b}. The consequence is a pronounced coherence in the flow field even without the influence of the magnetic field. \\
\TV{The application of a horizontal magnetic field supports an increase in the coherence of the flow pattern in a particularly striking way by transforming unsteady three-dimensional flows into stable two-dimensional structures. In this context, it is very interesting to point out that the application of small magnetic fields in the range of the three-dimensional flow $(Q < 10^4)$ does not show any measurable effects for the heat and momentum transport. From MHD turbulence it is known that the transition from isotropic to anisotropic turbulence starts at values of the magnetic interaction parameter $N= Q/Re \approx 1$ \citep{davidson1995, eckert2001}. When crossing this threshold the effect of the Lorentz force sets in, which prevents three-dimensional structures from absorbing the energy supplied by the thermal driving. Instead the development of quasi-two-dimensional structures are promoted. The interaction parameter reaches values of about 5 at the transition from the cell structure to the unstable roll regimes. The cell structure is completely three-dimensional and an amplification of the flow by the Lorentz force is not plausible in view of the described mechanism. Only with the emergence of the convection rolls are the quasi-two-dimensional structures available into which energy can be transferred. Accordingly, our measurements show a simultaneous increase of both momentum and heat transport in the regime of unstable roll structures.}\\

\TV{In conclusion,} we have shown how a stabilising, static magnetic field can significantly \TV{alter the flow dynamics such that the free-fall velocity is attained by the flow structure, resulting in enhanced} heat and momentum transport in thermal convection. \TV{These trends remained a consistent feature for all the Rayleigh numbers} investigated. \TV{It is likely that the optimum state} for the heat and momentum transport does not solely depend on $Q$, but also on a combination of $Q$, $Ra$ and $\Gamma$. \TV{Further investigation of this phenomenon with other combinations of parameters would therefore be desirable.}\\

\section*{Acknowledgements}
The authors thank Sanjay Singh, Megumi Akashi, Jonathan Aurnou, Susanne Horn, Takatoshi Yanagisawa, Yuji Tasaka and Sylvie Su for fruitful discussions. This work is supported by the Priority Programme SPP 1881 Turbulent Superstructures of the Deutsche Forschungsgemeinschaft (DFG) under the grant VO 2331/3. T.V. and F.S. also thank the DFG for the support under the grant VO 2331/1. The contribution of J.Y. in this project is financially supported by CSC (China Scholarship Council).

\bibliographystyle{jfm}


\end{document}